\documentclass[a4paper,fleqn,usenatbib]{mnras}

\usepackage[T1]{fontenc}
\usepackage{ae,aecompl}
\usepackage{graphicx}	% Including figure files
\usepackage{amsmath}	% Advanced maths commands
\usepackage{amssymb}	% Extra maths symbols

\title[White dwarf spin period variability in V2306 Cygni]{Detection of white dwarf spin period variability in the intermediate polar V2306 Cygni}

\author[V. V. Breus et al.]{
V. Breus,$^{1}$\thanks{E-mail: vitaly.breus@gmail.com (VB)}
K. Petr\'ik,$^{2}$
S. Zola$^{3,4}$
\\
$^{1}$Department of Mathematics, Physics and Astronomy, Odessa National Maritime University, Mechnikova 34, Odessa, Ukraine\\
$^{2}$M. R. Stefanik Observatory and Planetarium, Sladkovicova 41, Hlohovec, Slovak Republic\\
$^{3}$Astronomical observatory of the Jagiellonian University, ul. Orla 171, 30-244 Krakow, Poland\\
$^{4}$Mt. Suhora Observatory, Pedagogical University, ul. Podchorazych 2,  30-084 Krakow, Poland
}

\date{Accepted 2019 July 23. Received 2019 July 23; in original form 2019 May 09}

\pubyear{2019}

\begin{document}
\label{firstpage}
\pagerange{\pageref{firstpage}--\pageref{lastpage}}
\maketitle

\begin{abstract}
Magnetic cataclysmic variables are close binaries which consist of  
a compact object -- a white dwarf -- and a red dwarf filling its Roche Lobe. 
Such systems are physical laboratories which enable study of the influence 
of magnetic fields on matter flows. They often exhibit spin-up or 
spin-down of the white dwarf, while some systems exhibit more complex behaviour 
of the spin period change.
We monitor changes of the spin periods of white dwarfs in a sample of close 
binary systems to study interaction of the magnetic field and accretion processes 
as well as evolution of intermediate polars. 
Within the framework of our intermediate polar monitoring program, we obtained 
photometric CCD observations at several observatories. Two-period trigonometric 
polynomial fitting was used for determination of extrema timings. The (O-C) analysis 
was performed to study the variability of the orbital and spin periods of the systems.
Using data taken during 9 years of observations of the magnetic cataclysmic 
variable V2306 Cygni (formerly known as 1WGA J1958.2+3232), we detected the 
spin period variability which shows a spin-up of the white dwarf with 
a characteristic time of $(53\pm5)\cdot10^4$ years. The value of the spin 
period was $733.33976$ seconds with the formal accuracy of $0.00015$ seconds. 
We derived an improved value of the orbital period of the system 
to be $4.371523\pm0.000009$ hours.
\end{abstract}


%%%%%%%%%%%%%%%%%%%%%%%%%%%%%%%%%%%%%%%%%%%%%%%%%%

%%%%%%%%%%%%%%%%% BODY OF PAPER %%%%%%%%%%%%%%%%%%

\section{Introduction}

Intermediate polars, often called DQ Her type stars, are close binary systems 
consisting of a magnetic white dwarf and a main-sequence star filling 
the Roche lobe. The gravity of the primary component leads to capture of matter from 
the secondary component near the inner Lagrangian point. Due to the high angular 
momentum of the plasma leaving the Lagrangian point, the stream can not be accreted 
directly by the compact star, and instead it forms an accretion disk around the 
white dwarf, see e.g. \citet{LubowShu}. 

The magnetic field of the primary is strong enough to destroy the inner part of 
the accretion disk and matter is being accreted along the magnetic field lines, 
leading to the formation of one or two accretion columns near the magnetic poles. 
The matter forms a shock wave, heats up, and eventually settles down on the surface 
of the white dwarf. The accretion columns in such systems are often the brightest 
sources of polarized radiation in a wide spectral range from X-ray to radio. 
Rare outbursts have also been observed, e.g. \citet{hellier2004} 
and \citet{AndMultiple}.
% add here more proof of outbursting IP

Usually intermediate polars show two types of optical variability. The orbital 
variability is caused by the rotation of the system. During its revolution, 
we see the stars, an accretion disk and a hot spot at different angles. 
The orbital periods of intermediate polars are usually about 3--7 hours, 
see Table~1 and Fig.~11 from review paper by \citet{Patterson1994}. 
The spin variability is caused by the rotation of the white dwarf with one 
or two accretion columns, with the periods ranging from few to 
dozens of minutes. Therefore, the light curve is frequently a superposition 
of 2 different periodic variations and some aperiodic processes like flickering 
or outbursts, leading to changes between high and low luminosity states.

The pulsating X-ray source 1WGAJ1958.2+3232 was discovered by \citet{israel1998}. 
In 2003, it was assigned the variable star designation V2306 Cyg. The authors 
discussed the nature 
of pulsations with a period of 12 minutes and concluded that the source may be 
either a low-luminosity X-ray pulsar or an intermediate polar. \citet{negueruela2000} 
classified it as an intermediate polar based on its X-ray and optical properties. 
They noticed the double-peaked shape of the Balmer emission lines, indicating 
the presence of an accretion disk. \citet{uslenghi2000} suggested that the true 
white dwarf spin period could be twice that published earlier by \citet{israel1999} 
and mentioned that this object is one of the slowest rotators showing a 
double-peaked spin profile.

\citet{zharikov2001} derived an orbital period 
of $0^d.1802\pm0^d.0065$ (4h~36m) based on the photometric data and a somewhat higher value of $0^d.18152\pm0^d.00011$ 
from the radial velocity curve. They confirmed the ($733.82\pm1.25$)s white dwarf 
spin period both from spectroscopy and photometry. They argued for the presence 
of an accretion disk and the existence of a hot spot on it based on examination of the 
radial velocity curves, the light curve and the results of Doppler tomography.
\citet{uslenghi2001} reported variations of circular polarization with a period 
twice as long ($1467\pm25$) as the previously published white dwarf spin.

\citet{norton2002} suggested that the orbital period was ($5.387\pm0.006$) hours, 
however this value corresponds to the $-1$ day alias of the period value published 
by \citet{zharikov2001}. They also confirmed that the rotational period of the 
white dwarf is twice as long as the pulse period, indicating that two-pole 
accretion may be occurring in the system. \citet{zharikov2002}, using their own data along 
with those provided by A.~SNorton, repeated the analysis and derived a similar result 
to their previous determination of $0^d.181195\pm0^d.000339$. However, they 
mentioned that a longer time base of observations is needed to further improve 
the measured value of the period.

\citet{breus2015} analyzed the results of 6 years of photometric monitoring of V2306 Cyg. 
They were not able to find spin period variations from the (O-C) analysis, but found 
a regular cycle miscount for the previously published orbital period and a linear 
trend in the (O-C) diagram. They concluded that an orbital period of $0^d.181545\pm0^d.000003$ 
better fitted their data. The periodogram analysis revealed a period of $2.0183$ days, 
and they noticed significant changes of mean brightness level from one night 
to another. This was interpreted as a possible precession of the accretion disk in this system.

The paper is organized as follows: in the next Section we describe the photometric 
data gathered for V2306~Cyg and the instruments used, and Section~\ref{sec:OCAnalysis} 
presents the methods used for derivation of the high precision period values. 
In Sections~\ref{sec:OrbitalPeriod} and \ref{sec:SpinPeriod} we describe the results 
obtained for the orbital and spin periods detected in the light curves of V2306~Cyg. 
We discuss our findings in the last Section.

%-------------------------------------------------------------------

\section{Observations and data reduction}
\label{sec:Observations}

The first set of photometric data used in this research was obtained with 
the 60~cm ZeissCassegrain telescope at the Observatory and Planetarium in 
Hlohovec, Slovakia (ZC600). At this site two instruments were used: in the 
period 2012-2014 the SBIG ST-9 camera, while the ATIK 383L CCD was attached 
to this telescope in 2015-2018. In 2014-2016 we gathered photometric observations 
with the the 50~cm Cassegrain telescope at the Astronomical Observatory of 
the Jagiellonian University in Krakow, Poland (Zeiss50), equipped with an Andor 
DZ936 BV camera. Additionally, our analysis includes one run obtained in 2013 by 
O.~R.~Baransky with the 70-cm telescope of the Astronomical Observatory of Taras 
Shevchenko National University in Kiev, Ukraine (AZT-8). This telescope is 
equipped with a PL47-10 FLI camera and this run was previously analyzed by \citet{breus2015}.

The observation log is listed in Table~\ref{tabobs}, the total duration of each 
run is presented in hours. Typically we monitored the target over an entire night. 

Most of our data were taken with alternating wide band V and R filters.
Some observational runs were taken in BVR or R filters. 
Exposure times were between 30--60 seconds, adjusted according to weather conditions, 
instrument and filters used.
The overhead time, required for CCD readout and filter changes, was typically a 
few seconds. The reduction,s consisting of calibration of scientific images for 
bias, dark and flatfield and extraction of instrumental magnitudes, was carried 
out with the Muniwin software developed by \citet{Munipack}.
We used nearby comparison stars, listed in the AAVSO Variable Star 
Plotter\footnote{https://www.aavso.org/apps/vsp/} database. The chosen 
primary comparison star is No. 127 marked in the AAVSO chart for all data 
before 2014 and No. 137 for 2014 and later observations. The comparison stars 
were found to be constant in our entire data set. The final derivation of 
magnitudes was obtained using the multiple comparison stars method described 
by \citet{KimAndMC2004} and implemented in Multi-Column View by \citet{AndBak2004} 
(hereafter named MCV).

\begin{table}
	\centering
	\caption{The log of observations}
	\label{tabobs}
\begin{tabular}{ccccc}
\hline
From & To & Runs & Duration & Telescope \\
\hline
08.07.2012 & 16.07.2012 & 3 & 11.46 & ZC600 \\
27.07.2013 & 28.07.2013 & 2 & 9.04 & ZC600 \\
19.08.2013 & 19.08.2013 & 1 & 3.36 & AZT-8 \\
19.05.2014 & 19.05.2014 & 1 & 5.63 & Zeiss50 \\
09.06.2014 & 30.06.2014 & 4 & 16.94 & ZC600 \\
11.05.2015 & 01.06.2015 & 5 & 17.60 & Zeiss50 \\
08.06.2015 & 12.07.2015 & 6 & 20.05 & ZC600 \\
28.06.2016 & 10.07.2016 & 4 & 16.04 & Zeiss50 \\
24.06.2017 & 26.07.2017 & 6 & 17.31 & ZC600 \\
09.06.2018 & 30.07.2018 & 7 & 18.07 & ZC600 \\
\hline
\end{tabular}
Data were grouped by seasons. 
`Number of runs' means the number of nights when the object was observed and
total duration is given in hours.
\end{table}

Along with our own data we also included in the analysis all CCD observations 
spanning more than a few hours published in the AAVSO database. Barycentric corrections 
was applied to all geocentric Julian dates.

\section{O-C Analysis}
\label{sec:OCAnalysis}
To determine spin maxima and orbital minima timings we used the trigonometric 
polynomial approximation of the light curve implemented in MCV applied 
for each filter separately. We choose a 2-periodic model for smoothing in the form:

\begin{equation}
m(t)=m_0 + r_1\cos(\omega_1(t-T_{01})) + r_2\cos(\omega_2(t-T_{02}))
\label{twoperiodicfit}
\end{equation}
where $m(t)$ is the smoothed value of brightness at time $t$, $m_0$ is the 
average brightness of theoretical curve (generally different from the sample 
mean, see \cite{Andronov2003}. $\omega_j=2\pi/P_j$, $r_j$ is the semi-amplitude, and
$T_{0j}$ is the epoch for maxima of brightness of photometric wave with 
number $j$ and period $P_j$. This method has been previously widely used 
for approximation of observations of intermediate polars (see \cite{KimAnd2005}, 
\cite{exhya2013}).

As mentioned above, the orbital periods of intermediate polars are 
within the range between 3--7 hours. Due to the different length of nights across 
the year and atmospheric conditions, it was not always possible to cover the 
entire orbital period during one night. Only when an orbital minimum was well 
covered was it possible to use minima timings derived from trigonometric 
polynomial fits for the (O-C) analysis.

In the opposite case, when we measured only that part of the light curve between 
minimum and maximum, or in the case of short runs, we 
were unable to determine the timing from such data with adequate precision. 
Therefore, to proceed with the (O-C) analysis of orbital variability we derived 
a synthetic minimum time from the combined data taken over a few nights or one minimum 
per season. With this approach we significantly increased the accuracy of timings.
Spin periods of intermediate polars are short, usually between 12--20 minutes.
Therefore, we took one average spin maximum per night. The accuracy 
estimate of the value derived in this way is much better than for individual extrema.
Furthermore, to increase the accuracy of the spin maxima timing, in this analysis 
we have also merged a few nights.

Instead of the typical representation of the (O-C) diagram being a function of 
the extrema timings and the cycle number $E$, calculated from an ephemeris in the 
following form:

\begin{equation}
O-C=T-(T_0+P\cdot E),
\end{equation}

we have used the dependence of phase, i.e. $\phi=(O-C)/P$, with time or cycle number.
For a correct ephemeris, the phases should be concentrated near the zero value. 
A linear trend in the (O-C) diagram would correspond to a shift of observed extrema 
timings relative to the calculated values due to the inaccurate value of the period. 
A parabolic trend would indicate changes of period with time, while strictly periodic 
changes seen in the (O-C) diagram usually indicate the presence of a third body in the system.

%--------------------------------------------------------------------

\section{Improvement of the orbital period}
\label{sec:OrbitalPeriod}

\begin{figure}
	\includegraphics[width=\columnwidth,trim={1.6cm 0 1.6cm 0.5cm},clip]{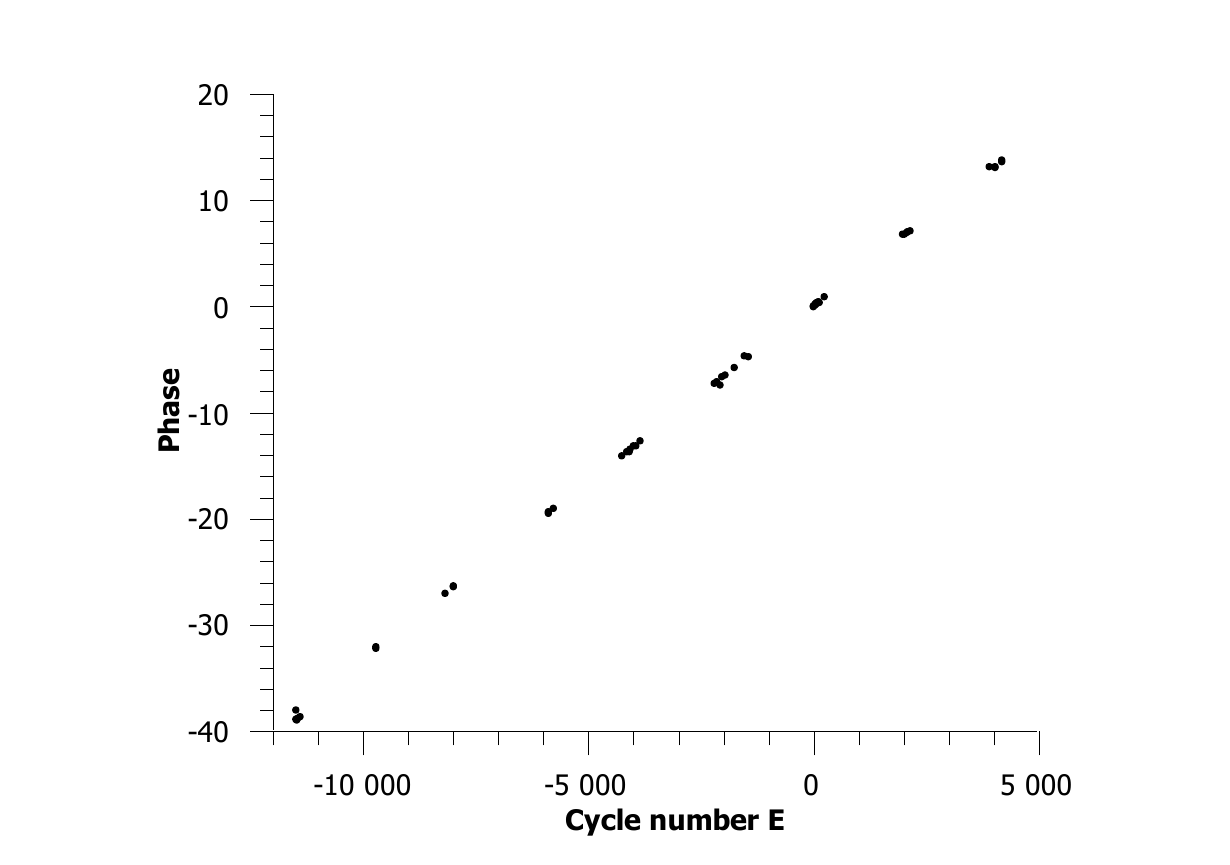}
    \caption{Dependence of phases of orbital minima on the cycle number -- 1st iteration.}
    \label{fig:fig_orboc1}
\end{figure}

\begin{figure}
	\includegraphics[width=\columnwidth,trim={1.6cm 0 1.6cm 0.5cm},clip]{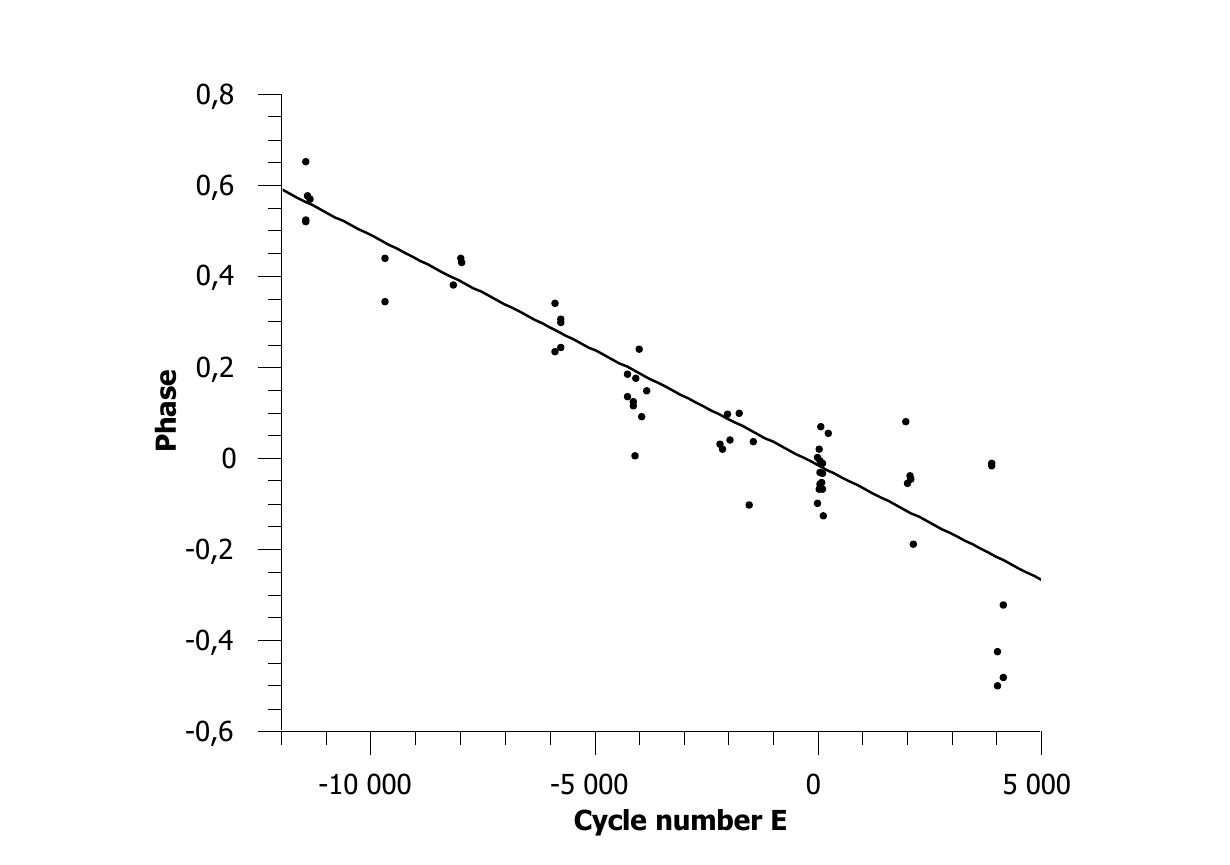}
    \caption{Dependence of phases of orbital minima on the cycle number -- 2nd iteration with weighted linear approximation.}
    \label{fig:fig_orboc2}
\end{figure}

Using all minima timings, we calculated the (O-C) diagram using an initial value 
of the orbital period of $P=0^d.181545$ (taken from the previous work by \citet{breus2015}) 
and an initial epoch of $T_0=2457570.27379$, determined for the night where the 
formal error estimate was smallest. It turned out that the cycle miscount was 
significant and required correction. Subsequently, we determined the statistically 
optimal linear weighted fit to the (O-C) (see Fig.~\ref{fig:fig_orboc1}). As a result,  
we obtained a more accurate value of the orbital period $0^d.182156\pm0.000001$.

With this new value, we repeated the calculation of orbital minima and the O-C 
analysis. The list of minima timings is available online.
%\ref{sec:online}.
%presented in Table~\ref{taborb}.
A second (O-C) diagram (see Fig.~\ref{fig:fig_orboc2}) has an amplitude of about 1 full 
period during all 9 years of monitoring and again we used a linear weighted fit to the (O-C). 
This resulted in an improved period value of $0^d.18214681\pm0.00000037$. After the third 
iteration the amplitude of the (O-C) diagram decreased to $0.4$ of P$_{orb}$, and most 
points fall in a band with width $\pm0.1$ P$_{orb}$ from the zero line.
We consider this orbital period value to be the most accurate and describing the analysed data 
best. The new determination has a formal statistical accuracy about 3 orders of magnitude 
smaller then that determined by \citet{zharikov2002} and a order of magnitude smaller 
then our previous result (see \citet{breus2015}). We found no periodic variations of 
the orbital period.

%--------------------------------------------------------------------

\section{Spin period variability}
\label{sec:SpinPeriod}

\begin{figure}
	\includegraphics[width=\columnwidth,trim={1.6cm 0 1.6cm 0.5cm},clip]{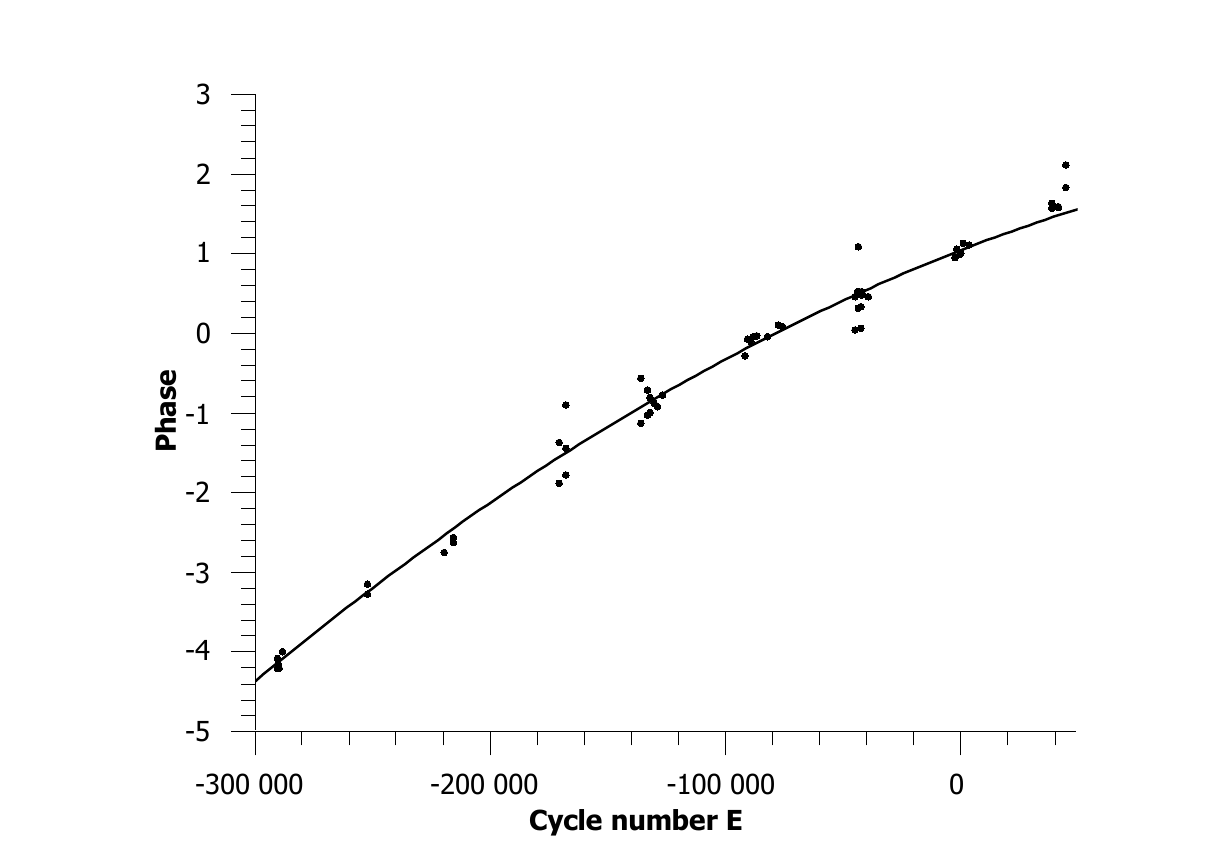}
    \caption{Dependence of phases of spin maxima on the cycle number with weighted parabolic approximation.}
    \label{fig:fig_spinoc1}
\end{figure}

Due to the similar shape of humps seen in the light curve, we used half of the 
spin period $0^d.008487595$ and an initial epoch of $2457950.46628$ to determine 
spin maxima timings (a table is available in the online version).
%(see Table~\ref{tabspin}).
The (O-C) diagram calculated 
for these data shows a regular cycle miscount of about $0.8$ cycles per year. 
We introduced a correction to account for this miscount, presented in Fig.~\ref{fig:fig_spinoc1}, 
and performed computations to derive best fits to the (O-C) diagram. 
We consider a cubic fit not justified, as the coefficient $Q_3$ is about $1.5\sigma$ 
of its error estimate. 
Our next trial was a quadratic weighted polynomial fit in the form: 
\begin{equation}
%\phi= -0.29(2) + 0.0000157(2)~(E-E_0) - 0.000000000022(2)~(E-E_0)^2
\phi= -0.29(2) + 0.0000157(2)\cdot(E-E_0) - 22(2)\cdot10^{-12}\cdot(E-E_0)^2
\end{equation}
Here $E_0=-97767$ is an integer cycle number close to mean time of observations and the 
quadratic term reaches $11\sigma$. This fit may correspond to the current value of the 
half of the spin period of $0^d.0084877287\pm0.0000000017$ and characteristic 
time of change of spin period:
  
\begin{equation}
\tau=P_0/|dP/dt|=P_0^2/|2Q_2|=(53\pm5)\cdot10^4 $ years$
\end{equation}

%--------------------------------------------------------------------

\section{Conclusions}
\label{sec:Conclusions}

From the analysis of observations gathered during 9 years of photometric monitoring 
of the intermediate polar V2306 Cygni, we discovered its white dwarf spin period 
variability. The characteristic spin-up time is $(53\pm5)\cdot10^4$ years and this 
value is the same both for the spin period and the twice shorter pulse period. 
The value of $dP/dt = (-8.73\pm0.79)\cdot10^{-11}$ for the spin period (twice 
larger than the value obtained for the pulse period) has an order typical of 
intermediate polars.

For the epoch of 2017, we derived the white dwarf rotation rate  value 
of $1466.6795$ seconds, with a formal accuracy of $0.0003$ seconds. This is 
consistent with that determined by \citet{norton2002} ($1466.66\pm0.06$ seconds), 
however, our determination based on a longer span of data is more accurate. 
From the (O-C) analysis we improved the value of the orbital period of V2306 Cyg 
resulting in a value of $P=4.371523\pm0.000009$ hours.

Spin period variations are frequently observed in intermediate polars and are 
typically detectable on a timescale of decades (see e.g. \citet{andaspc2017}). 
Some intermediate polars show a constant decrease (e.g. EX Hya, \citet{exhya2013}, 
BG CMi, \citet{KimAndMC2004}), while others exhibit an increase of their white 
dwarfs' rotation rates (PQ Gem, V1223 Sgr, AE Aqr mentioned by \citet{norton2002}). 
Even more complicated period changes, an increase followed by a decrease of rotation 
periods, were detected during observations covering a couple of decades 
(e.g. FO~Aqr, \citet{breus2012fo} and V405~Aur, Breus et al. (in preparation)). 
On the other hand, V1323 Her does not show a statistically significant spin period 
change (see \citet{petrikaspc}). 

\citet{Miguel2133} compiled published spin-up rates for intermediate polars. Among 
16 objects listed there, V2306 Cyg has one of the longest orbital and spin periods. 
The value of $dP/dt$ for this object, along with orbital and spin periods, resemble 
the properties of FO~Aqr - "The King of Intermediate Polars", which exhibited at 
least three low states during recent years (\citet{kennedy2017}, \citet{lit2018}).

The variations of the white dwarf rotation rates in intermediate polars can be caused 
by a change of angular momentum of the white dwarf due to accretion,  precession of the 
rotation axis of the magnetic white dwarf (see \citet{Andronov2005} and 
\citet{Tovmassian2007}) or the presence of a third body in the system.

%Referee's comment:
%accretion of matter takes place in all intermediate polars. Why in this particular system it results in spin-up?
%Please clarify, or abstain from making such claims. 

%Block removed
%The detected spin-up of the white dwarf in V2306 Cygni likely may be caused 
%by an increase of its angular momentum due to accretion of matter.

%-------------------------------------------------------------------

\section*{Acknowledgements}
	We acknowledge with thanks the variable star observations from the AAVSO 
International Database (https://www.aavso.org/) contributed by observers worldwide 
and used in this research and personally William Goff, James Jones, Joseph Ulowetz, 
David Boyd, Michael Cook, Richard Sabo, Donald Collins, Lewis Cook and Etienne Morelle. 
We thank prof. Ivan L. Andronov for helpful discussions on this paper, Olexander 
Baransky for providing his observations and Greg Stachowski for language corrections. 

%%%%%%%%%%%%%%%%%%%%%%%%%%%%%%%%%%%%%%%%%%%%%%%%%%



%\bibliographystyle{mnras}


%%%%%%%%%%%%%%%%%%%%%%%%%%%%%%%%%%%%%%%%%%%%%%%%%%

%%%%%%%%%%%%%%%%% APPENDICES %%%%%%%%%%%%%%%%%%%%%
%\appendix
%If you want to present additional material which would interrupt the flow of the main %paper,

\appendix
\section{Supporting Information}
\label{sec:online}
Supplementary data are available online at \url{http://uavso.org.ua/data/v2306cyg/}.
\textbf{v2306cyg-orbital.txt} contains the list of orbital minima timings.
\textbf{v2306cyg-spin.txt} contains the list of spin maxima timings.
Both are tab-separated text files, the columns are BJD, error estimate, telescope and AAVSO observer code (see Section~\ref{sec:Observations}).

%%%%%%%%%%%%%%%%%%%%%%%%%%%%%%%%%%%%%%%%%%%%%%%%%%

% Don't change these lines
\bsp	% typesetting comment
\label{lastpage}
\end{document}